\documentclass[conference]{IEEEtran}
\IEEEoverridecommandlockouts
\usepackage{graphicx,cite,calc,xcolor,subfigmat,amssymb,amsmath,mathrsfs,dsfont,hyperref,epstopdf}
\usepackage[normalem]{ulem}
\usepackage{algorithm}
\usepackage[switch]{lineno}

\usepackage{algpseudocode}

\usepackage{dsfont}
\usepackage{bbm}
\usepackage{soul}
\usepackage{xcolor}
\usepackage{amsmath}
\newcommand{\dsm}{\mathbbm{m}}
\newcommand{\dsh}{\mathds{h}}
\usepackage{float}
\usepackage[T1]{fontenc}
\usepackage{soul} 
\usepackage{xcolor}
\usepackage{optidef}
\usepackage{listings}
\usepackage{enumitem}
\usepackage{hyperref}
\makeatletter
\newcommand*{\rom}[1]{\expandafter\@slowromancap\romannumeral #1@}
\makeatother
\usepackage {amsmath}
\definecolor{highlightcolor}{RGB}{255,255,0}

\usepackage{newtxtext,newtxmath}
\title{Energy Sustainability in Dense Radio Access Networks via High Altitude Platform Stations}

\begin{document}

 \author{\rm{Maryam Salamatmoghadasi}, \rm{Amir Mehrabian}, \rm{Halim Yanikomeroglu}
 \thanks{All the authors are with the Department of Systems and Computer Engineering, Carleton University, Ottawa, ON K1S5B6, Canada.; email: \texttt{maryamsalamatmoghad@cmail.carleton.ca, amirmehrabian@cunet.carleton.ca, halim@sce.carleton.ca }.}}

\maketitle
\begin{abstract}
The growing demand for radio access networks (RANs) driven by advanced wireless technology and the ever-increasing mobile traffic, faces significant energy consumption challenges that threaten sustainability. To address this, an architecture referring to the vertical heterogeneous network (vHetNet) has recently been proposed. Our study seeks to enhance network operations in terms of energy efficiency and sustainability by examining a vHetNet configuration, comprising a high altitude platform station (HAPS) acting as a super macro base station (SMBS), along with a macro base station (MBS) and a set of small base stations (SBSs) in a densely populated area. By intelligently managing SBSs’ sleep mode and employing HAPS’s potentials and additional capacity, our approach aims to minimize vHetNet energy consumption.
The proposed method dynamically determines which SBSs to switch off based on the traffic load of SBSs, MBS, and HAPS. This innovative approach offers a flexible and promising solution to enhance network sustainability, energy efficiency, and capacity utilization without compromising the user quality-of-service (QoS). We show that our proposed method offers a scalable solution with comparable performance to exhaustive search (ES) as the optimal solution in terms of energy efficiency. Furthermore, incorporating HAPS, significantly improves grid power consumption, compared to having no offloading, reducing it by 30\% for a large number of SBSs.
\end{abstract}
\section{Introduction}
The proliferation of conventional communication services, particularly those relying on multimedia content, along with the widespread adoption of internet of things (IoT) communication in various domains, has led to an exponential surge in mobile traffic demand in radio access networks (RANs). To meet this growing demand, there is a need for RAN densification. Moreover, the rapid growth in the number of connected devices is expected to drive 6G networks to support approximately 10 million devices per square kilometer, representing a tenfold increase compared to the connection density requirements of 5G \cite{9773096}.
\\
\indent In the face of this unprecedented growth, RANs confront a complex and profound challenge: the escalating energy consumption that poses a threat to both environmental sustainability and economic feasibility. It is imperative to recognize the critical importance of mitigating this energy consumption challenge. Base stations (BSs) are the major energy consumers,
accounting for about 60\%–80\% of the total energy consumption
in cellular networks \cite{3333333}. Also, the overall power requirements of 5G networks are estimated to be much higher than those of 4G due to the densification of
the network and the increased energy usage of BSs and user
equipment \cite{2222222}, potentially surpassing 10 times the energy demand. Given the temporal and spatial fluctuations in the traffic load of cellular networks, the conventional approach of keeping all BSs, particularly small base stations (SBSs), in an always-on state, even when they are not serving any users, leads to energy inefficiency. Consequently, there has been considerable research interest in load-adaptive network operation, which involves powering off or operating SBSs in low power modes during periods of minimal or no traffic, aiming to conserve energy \cite{3333333},\cite{7446253}, and \cite{4444444}. 
In the conventional heterogeneous network (HetNet) architecture, certain elements of SBSs must remain operational even in sleep mode due to maintaining quality-of-service (QoS) experienced by users\cite{9344664}.\\
\indent Handling the aforementioned challenges of SBSs switching off in the conventional HetNet, innovative resource on demand (RoD), and cell switching off approaches have been introduced to minimize network energy consumption while reducing the necessity using convex optimization and machine learning \cite{9344664}, \cite{ABUBAKAR2022101643}. In this context, the strategic integration of non-terrestrial networks (NTN), known as vHetNet, presents significant potential as an effective solution to enhance capacity precisely in high-demand areas and times \cite{9583591}. Notably, this integration offers the advantage of alleviating the need for additional energy consumption from the power grid, making it a compelling option for optimizing network performance and sustainability. Specifically, the integration of a high altitude platform station (HAPS) into the network infrastructure presents a noteworthy solution. Positioned in the stratosphere, a HAPS equipped with super macro base station (SMBS) holds the capability to offer significant supplementary capacity across a vast geographical region \cite{9380673}. 
In \cite{song2023high}, the offloading of the entire MBS load to HAPS is discussed while in\cite{10001388}, cell switching is examined using an exhaustive search (ES) approach.
However, the integration of HAPS in densely populated areas with terrestrial networks, with a focus on flexibility and sustainability, remains an understudied aspect in existing literature \cite{9773096},\cite{7925666}. \\
\indent Motivated by the potentials of HAPS, this study aims to leverage the self-sustainability of NTN by offloading part of terrestrial traffic demand in a vHetNet. By incorporating the use of solar energy for HAPS, we envision an efficient solution with sustainable operational expenditure\cite{7925666}.
To the best of our knowledge, this is the first work that investigates the detailed benefits of HAPS in minimizing vHetNet power consumption within a macro cell. 
As the main contribution of this paper, we investigate the sustainability of a vHetNet including one HAPS, one MBS, and several SBSs. We formulate an optimization problem within this scenario and introduce a scalable traffic-aware cell switching and traffic offloading algorithm to control the ON/OFF status of SBSs. Notably, our proposed algorithm achieves a performance very close to the optimal solution provided by ES approach while harnessing the potential of HAPS. This algorithm contributes significantly to reducing vHetNet power consumption without compromising the QoS. We demonstrate the effectiveness of our proposed method through comprehensive simulations and performance comparisons with baseline schemes.

\section{System Model}

\subsection{Network Model}

The network model considered in this study is depicted in Figure \ref{fig-2}. This is a vHetNet comprising a dense configuration which incorporates SBSs within the coverage area of a MBS and a HAPS. 
SBSs have the role of providing data services, 
whereas the MBS and HAPS are responsible for maintaining consistent coverage, managing control signals, as well as providing data services. 
Additionally, HAPS plays a crucial role in coordinating offloading traffic of SBSs within its coverage area. This task involves monitoring the traffic loads of SBSs and making informed decisions about which ones to switch off during periods of low traffic intensity, while considering the available capacity of the MBS and HAPS.

\subsection{Network Power Consumption}
Based on the energy-aware radio and network technologies (EARTH) power consumption model \cite{6056691}, the power consumption by the $i$-th BS at any given time, denoted as $ {P_i} $, is given by \cite{7925662}
\begin{equation}
P_i  = \left\{ {\begin{array}{*{20}c}
   {\begin{array}{*{20}c}
   {P_{O,i}  + \eta _i \lambda _i P_{T,i} }, & {0 < \lambda _i  < 1,}  
\end{array}}  \\
   {\begin{array}{*{20}c}
   {P_{S,i} },  \; \;   \; \; \; \; \; \; \;\; \; \; \; \; \; \; \; \; \; \; \; \; \; & {\lambda _i  = 0},  \\
\end{array}}  \\
\end{array}} \right.
\label{eq1}
\end{equation}
where $ {P}_{O,i} $ corresponds to the operational circuit power consumption, $ {\eta _i} $ is the power amplifier (PA) efficiency, $ {\lambda _i} $ is the load factor, $ {P}_{T,i} $ is is the transmit power, and $ {P}_{S,i} $ is the sleep circuit power consumption. 
The network's instantaneous power consumption, denoted as $ {P} $, is calculated as follows
\begin{equation}
P = P_H  + P_M  + \sum\limits_{i = 1}^s {P_i }, 
\label{eq2}
\end{equation}
where $ {P_H} $ and $ {P_M} $ are the HAPS and MBS power consumption at a given moment, respectively, and $ {s} $ is the number of SBSs in the deployed network.
\section{Problem Formulation}
Here, we are attempting to figure out the most efficient energy saving state that provides the required QoS to users. Defining a state vector ${\rm \Delta = \{ 
\delta _1 ,\delta _2 ,...,\delta _s \} } $ involves determining which SBSs are ON at a given time $ {t} $. ${\rm \delta _i  \in \left\{ {0,1} \right\}}$ shows the current state of SBS, where 0 indicates OFF, and 1 indicates ON. Here, the MBS and HAPS are always ON and   $\delta _M = \delta _H = 1$. 
For a SBS, when ${\rm \delta _i }$ changes from 1 to 0 at time $ {t} $, either the MBS or HAPS (based on the SBS location) could allocate its traffic
\begin{equation}
\lambda _{k,t}  = \lambda _{k,t - 1}  + \phi _{i,k} \lambda _{i,t - 1}, \; \; k=M \;  \mathrm{or}\;   H, 
\label{eq3}
\end{equation}
\begin{equation}
\lambda _{i,t}  = 0,
\label{eq5}
\end{equation}
where $ {\rm \lambda}_{k,t} $ is the load applied to MBS or HAPS, while $ {\rm \lambda}_{i,t} $ is the load applied to the $i$-th ${\rm SBS}$. Additionally, ${\phi_{i,k}}$ is the relative capacity of $i$-th ${\rm SBS}$ with respect to the MBS or HAPS, such that ${\phi _{i,k}  = {{C_i } \mathord{\left/
 {\vphantom {{C_i } {C_k }}} \right.
 \kern-\nulldelimiterspace} {C_k }}}$, ${k=M}$ or ${H}$, where $ {C_i} $ represents the total capacity of $i$-th ${\rm SBS}$, while $ {C_k} $ is related to the total capacity of MBS or HAPS.
On the other hand, when ${\rm \delta _i }$ changes from 0 to 1, the MBS or HAPS offloads some traffic to the $i$-th ${\rm SBS}$, such that ${\lambda _{i,t}  = {{\tau _i } \mathord{\left/
 {\vphantom {{\tau _i } {C_i }}} \right.
 \kern-\nulldelimiterspace} {C_i }}}$ and
\begin{equation}
 \lambda _{k,t}  = \lambda _{k,t - 1}  - \phi _{i,k} \lambda _{i,t},  \; \; k=M \; \mathrm{or}\;   H, 
 \label{eq9}
\end{equation}
where ${\tau _i }$ refers to the traffic of $i$-th ${\rm SBS}$.
Accordingly, the problem can be expressed as follows

\begin{mini!}
		{\Delta}{P(\Delta )}
		{}{}{}
		\addConstraint{\lambda _H}{\le 1}
 \addConstraint{\lambda _M}{\le 1}
  \addConstraint{{\delta _i  \in \{ 0,1\}, }}  {{\; \;i = 1,...,s}},
	\end{mini!}
where
\begin{equation}
\begin{aligned}
P(\Delta )&={(P_{O,H}  + \eta _H \lambda _H } P_{T,H} )+{(P_{O,M}  + \eta _M \lambda _M } P_{T,M})\\
 &\hspace{-2em}\;\;\;\;\;\;\;+(\sum\limits_{i = 1}^s {(P_{O,i}  + \eta _i \lambda _i  P_{T,i} )\delta _i +P_{S,i}(1-\delta _i))}.
\end{aligned}
\end{equation}

The constraints in the optimization problem guarantee that the capacity of the MBS and HAPS are not exceeded, thereby satisfying the QoS requirement. However, solving this binary optimization problem becomes challenging due to its non-convex nature and the increased complexity when dealing with a large $s$. To address this, we propose a suboptimal and straightforward solution.
\begin{figure}[t]
\centerline{\includegraphics[width = 7.cm ]{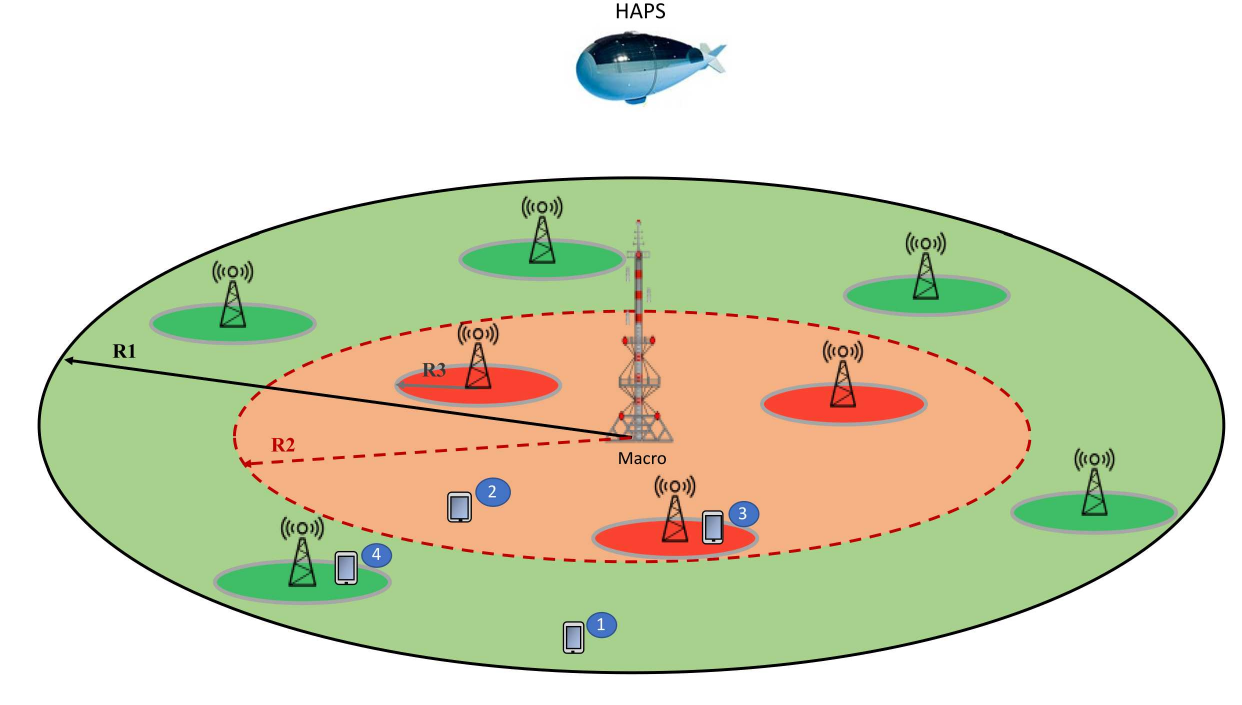}}
\caption{HAPS+MBS network model.}
\label{fig-2}
\vspace{-1em}
\end{figure}
  \section{Proposed Solution}
In this paper, our aim is to identify the best state vector, that determines which SBSs should be switched ON/OFF at different times to optimize the network energy consumption. Regarding possible network configurations, the complexity of ES method grows exponentially, particularly with an increasing number of SBSs. For instance, with $s=30$, ES should explore a vast search space with $2^{30}$ different options for $\delta_i$s. In contrast, our solution is computationally efficient. It achieves a performance that closely approximates ES approach by sorting SBSs in ascending order based on their load factors and intelligently offloading their traffic to the MBS or HAPS. 
This solution takes into account two limitations. Firstly, the QoS of users should be satisfied by ensuring that the MBS and HAPS load factors are less than 1, and they are able to serve the offloaded traffic. Secondly, to achieve an optimal state vector for switching any SBSs ON/OFF, a threshold on the load factor needs to be established.  
For instance, if changing the state of a SBS from 1 to 0 would result in power consumption savings for the network, the difference in power consumption (${\Delta P}$) between the two consecutive times should be negative
\begin{equation}
\Delta P = P_t  - P_{t - 1}  < 0.
\label{eq12}
\end{equation}
If we assume that the load of the $j$-th ${\rm SBS}$ can be offloaded to the HAPS based on its location and the available capacity of the HAPS, \eqref{eq12} can be expanded as
\begin{equation}
\begin{array}{l}
 \Delta {P = P}_{{H,t}} + {P}_{{M,t}}  + {P}_{{j,t}}  + \sum\limits_{i = 1,i \ne j}^s {{P}_{{i,t}} }  \\ 
  \;\;\;\;\;\;- ({P}_{{H,t - 1}} +  {P}_{{M,t - 1}}  + {P}_{{j,t - 1}}  + \sum\limits_{i = 1,i \ne j}^s {{P}_{{i,t - 1}} } ). \\ 
 \end{array}
 \label{eq13}
\end{equation}
Since we have assumed that only the state of the $j$-th ${\rm SBS}$ is changed, while the states of all other SBSs and the MBS are maintained, \eqref{eq13} can be expressed as
\begin{equation}
\Delta P = P_{H,t}  + P_{j,t}  - (P_{H,t - 1}  + P_{j,t - 1}). 
\label{eq14}
\end{equation}
Thus, using \eqref{eq1}, \eqref{eq3}, and \eqref{eq5}, \eqref{eq14} can be rewritten as 
\begin{equation}
\begin{array}{l}
 \Delta P = P_{O,H}  + \eta _H (\lambda _{H,t - 1}  + \phi _{j,H} \lambda _{j,t - 1} )P_{T,H}  + P_{S,j}  \\ 
  \;\;-(P_{O,H}  + \eta _H \lambda _{H,t - 1} P_{T,H}  + P_{O,j}  + \eta _j \lambda _{j,t - 1} P_{T,j} ). \\ 
 \end{array}
 \label{eq15}
\end{equation}
Then, we can simplify \eqref{eq15} to
\begin{equation}
\Delta P = \eta _H \phi _{j,H} \lambda _{j,t - 1} P_{T,H}  + P_{S,j}  - P_{O,j}  + \eta _j \lambda _{j,t - 1} P_{T,j}. 
\label{eq16}
\end{equation}
As a result, in order to satisfy \eqref{eq12} and save some network power consumption, the following limitation on the $j$-th ${\rm SBS}$ load factor is required
\begin{equation}
\lambda _{j,t - 1}  < \frac{{P_{O,j}  - P_{S,j} }}{{\phi _{j,H} \eta _H P_{T,H}  - \eta _j P_{T,j} }}.
\label{eq17}
\end{equation}
 The right hand side of \eqref{eq17} is defined as the threshold for HAPS offloading, ${\rm \rho _H}$. Similarly, there exists another threshold, denoted as ${\rm \rho _M}$, for offloading decisions related to the MBS. It is important to note that to satisfy \eqref{eq17} and make the load factor of SBSs a contributing factor for offloading decisions, two conditions must be met. Firstly, since the load factor is always in the range of 0 to 1, the defined threshold should also be set below 1. Secondly,  the denominator of the fraction should be positive, which imposes a constraint on $\phi _{j,H}$. Therefore, the selection of an appropriate ratio between the capacity of SBSs and the MBS or HAPS is crucial to optimize the system performance. 
 Another noteworthy observation regarding \eqref{eq17} is that the optimal state vector allows for more SBSs to be switched off when they are lightly loaded, thereby facilitating the fulfillment of the offloading condition in \eqref{eq17}. In contrast, SBSs with higher loads are assumed to remain active to ensure optimal performance.

\begin{algorithm}[t]
\caption{Proposed HAPS+MBS offloading}
\begin{algorithmic}[1]
\Procedure{Power Consumption Minimization}{}
\State \textbf{Input:} Traffic loads of HAPS, MBS, and SBSs
\For{all time interval $\Delta T_t$}
    \State sort cells  i $\in$  $\dsh$ based on $\lambda _i$ in ascending order:
      \\
      \;\;\;\;\;\;\;\;\;\; $\lambda _0  < \lambda _1  < ... < \lambda _j  < ... < \lambda _{\left| \dsh\right|}$
\\
 \;\;\;\;\;\;\;\;\;\;offload from the SBS with the lowest load factor\\ 
 \;\;\;\;\;\;\;\;\;\;until we reach:\\
 \;\;\;\;\;\;\;\;\;\; $\lambda _H  = 1$ or $\lambda _i  > \rho _H$ 
\\
 \;\;\;\;\;\;\;\;\;\;sort cells  i $\in$  $\dsm$ based on $\lambda _i$ in ascending order:
\\
 \;\;\;\;\;\;\;\;\;\;$\lambda _0  < \lambda _1  < ... < \lambda _p  < ... < \lambda _{\left| \dsm\right|}$
\\
 \;\;\;\;\;\;\;\;\;\;offload from the SBS with the lowest load factor \\
 \;\;\;\;\;\;\;\;\;\;until we reach:\\
 \;\;\;\;\;\;\;\;\;\; $\lambda _M  = 1$ or $\lambda _i  > \rho _M$ 
\EndFor
\State \textbf{Output:} ${P_{min}(\Delta )}$
\EndProcedure
\end{algorithmic}
\label{alg1}
\end{algorithm}
In Algorithm \ref{alg1}, $\dsh$ and $\dsm$ are sets of SBSs which their traffic could be offloaded to HAPS and MBS, respectively.
It is important to emphasize that, unlike the approach in [6], our algorithm does not rely on iterative steps.
Figure \ref{fig-2} illustrates the schematic performance of our proposed algorithm, highlighting the capabilities of each server in our vHetNet. the HAPS serving area, which includes MBS cell edge areas, is represented in pale green with radius of ${R1}$ (user 1), while the MBS coverage area is depicted in pale red with radius of ${R2}$ (user 2). To enhance service reliability, multiple SBSs are deployed within both the MBS and HAPS serving areas with radius of ${R3}$. In our proposed approach, we leverage prior knowledge about the locations of SBSs and their serving area signal coverage and quality as received by the MBS and HAPS. This enables us to create an architecture-aware vHetNet that facilitates efficient cell switching and traffic offloading. Consequently, the serving areas supported by SBSs that meet the offloading condition are divided into two segments. Firstly, SBSs located within the MBS coverage area are indicated in dark red, such as the cell containing user 3. Secondly, SBSs situated in areas with poor MBS coverage due to long distances, thus relying on the HAPS as a backup server for offloading, are depicted in dark green, exemplified by the cell where user 4 is located.
To integrate this information on the location of SBSs in the offloading process, we introduce a priority factor denoted as ${\gamma}$. This factor is defined as the ratio of the number of SBSs that can be offloaded to the HAPS (dark green SBSs) to the total number of SBSs in the network. A higher priority factor indicates a larger number of SBSs that can benefit from HAPS support. In such cases, the HAPS plays a crucial role in effectively managing cell switching and traffic offloading. 

\section{Performance Evaluation}
Our network is built by implementing the proposed cell switching framework with parameters specified in Table \ref{table:nonl}. Table \ref{table:non2} presents the power consumption characteristics of the BSs used in the simulations. We assume that the MBS and HAPS have the same circuit power consumption but different PA efficiency. The rationale behind considering different PA efficiency values is to meet the threshold limitation described in \eqref{eq17}. It is worth noting that the consideration of higher PA efficiency for HAPS is a pessimistic assumption as it results in increased power consumption. We use the SoftMax function to distribute traffic among the SBSs randomly and the remaining traffic is distributed between the MBS and HAPS.
\begin{centering}
\begin{table}[t] 
\caption{SIMULATION PARAMETERS}
\centering 
\begin{tabular}{|c|c|} 
  \hline 
Parameters&Value
\\ 
\hline
Number of time slots&500
\\
HAPS cell radius, R1 & 564 m
\\
Macro cell radius, R2 & 471 m
\\
Small cell radius, R3 & 60 m
\\
Priority factor, $\gamma$ & 0.7
\\
HAPS capacity&40 Gbps
\\
MBS capacity&10 Gbps
\\
SBS capacity&5 Gbps
\\
\hline
    \end{tabular} \label{table:nonl} 
     \vspace{-2em}
    \end{table}
    \end{centering}
   \subsection{Evaluation of Benchmarks}
For comparison purposes, there are different benchmark methods to be utilized, which are as follows:
\subsubsection{No Offloading Method} In this method, all SBSs remain ON at all times. As a result, QoS is not a consideration in this method, since all users are served by the same server that they were originally associated with.
\subsubsection{MBS Offloading Method} Using this algorithm, SBSs with load factors below than the defined threshold are sorted in ascending order. SBSs are then switched off when there is no more capacity available at MBS, while the remaining SBSs are kept on. HAPS is also available through this method, but no offloading is performed from SBSs to HAPS (${\gamma=0}$).
\subsubsection{Exhaustive Search} This method is a comprehensive approach that explores all possible state vectors of SBSs' OFF/ON states to find the optimal configuration. It aims to ensure user QoS by selecting the option with the least energy consumption, considering the availability of HAPS and MBS.
Due to its high computational complexity, we limited the execution of this method to scenarios involving a small $s$.
 
\begin{centering}
\begin{table}[b] 

\caption{DIFFERENT BSs POWER PROFILE}
\centering 
\begin{tabular}{|cc|ccc|} 
  \hline 
& & &Power Consumption& \\
BS Type&Efficiency&Transmit&Operational&Sleep
\\ 
&$\eta _i$&$P_{T,i}$$[W]$&$P_{O,i}$$[W]$&$P_{S,i}$$[W]$
\\  
\hline
HAPS&15&20&130&75
\\
MBS\cite{6056691}&4.7&20&130&75
\\
SBS\cite{6056691}&2.6&6.3&56&39 
\\[1ex] 
    \hline 
    \end{tabular} \label{table:non2} 
    \vspace{-1em}
    \end{table}
\end{centering}
\subsection{Key Performance Indicators}
The indicators for evaluating the performance of the proposed solution are:
\subsubsection{Energy Efficiency} This indicator is defined as the ratio between the total traffic demand and the total power consumption of the network in the area under consideration.
\subsubsection{Network Power Consumption} The total power consumption of the network using \eqref{eq2} is an effective metric for evaluating the performance and particularly the sustainability of different methods.
\subsubsection{Grid Power Consumption} Grid power consumption evaluation excludes HAPS' solar power contribution, offering valuable insights into the network's energy usage from grid sources \cite{9773096}.
\subsubsection{Capacity Utilization} This indicator is defined as the ratio of the served traffic demand by the sum active RAN capacity of the network \cite{7925666}.
\begin{figure}[t]
\centerline{\includegraphics[width = 7.cm ]{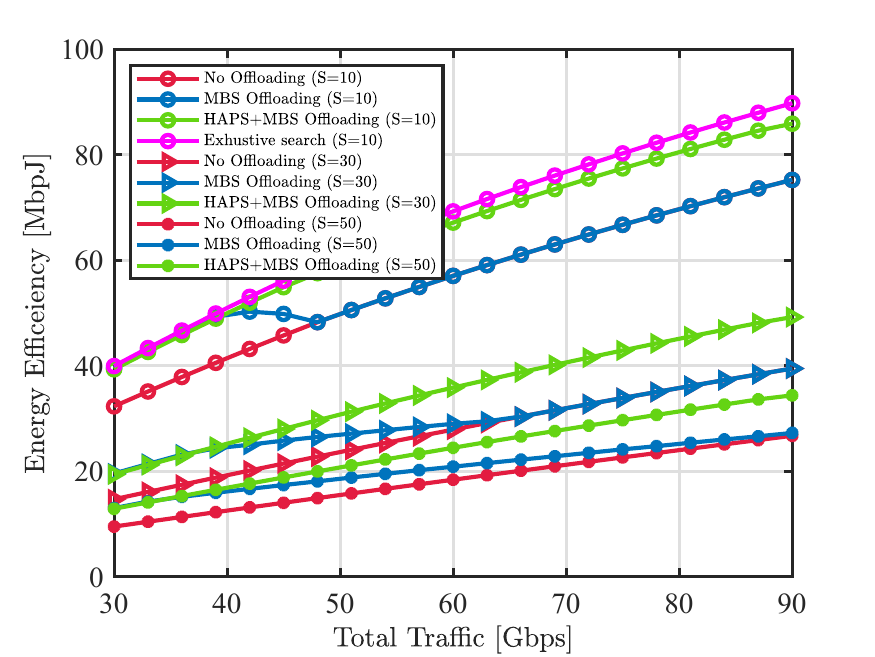}}
\caption{Energy efficiency vs. total traffic for different number of SBSs (${s=10, 30, 50}$).}
\label{fig4}
  \vspace{-1em}
\end{figure}
\subsection{Simulation Results}
\begin{figure}[t]
\centerline{\includegraphics[width = 7.cm ]{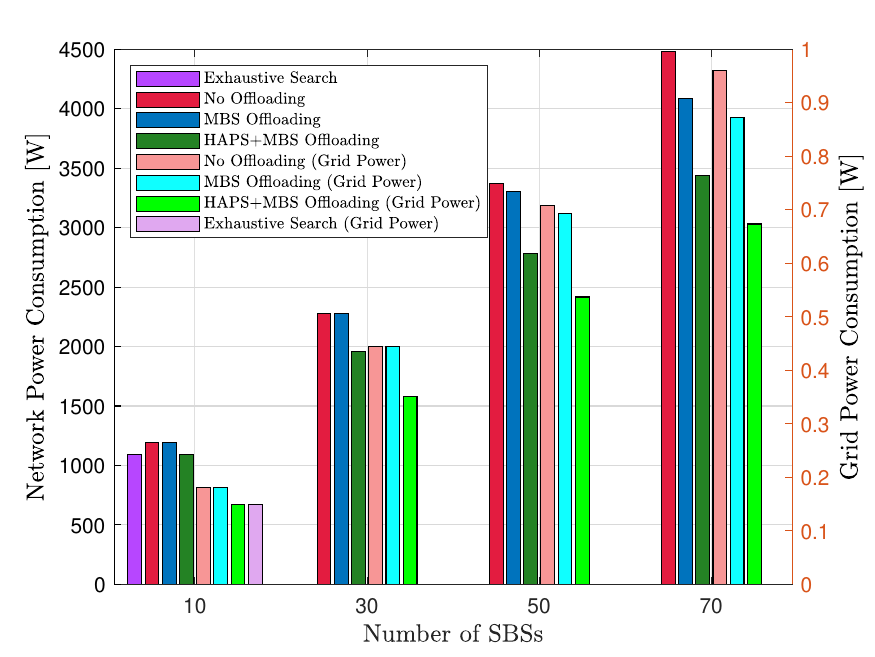}}
\caption{Network power consumption vs. number of SBSs.}
\label{fig3}
  \vspace{-1em}
\end{figure}

\begin{figure}[t]
\centerline{\includegraphics[width = 7.cm ]{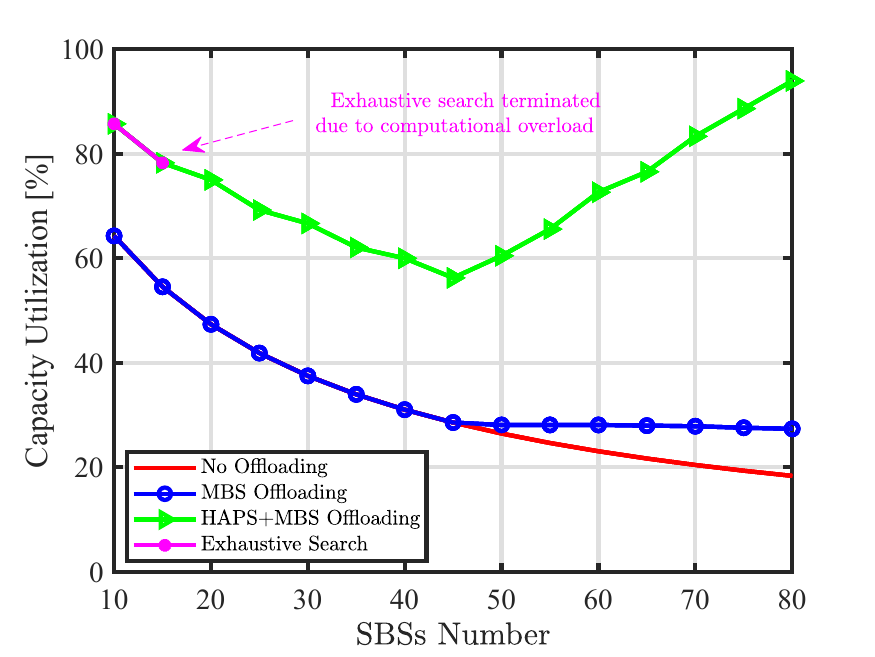}}
\caption{Capacity utilization vs. number of SBSs.}
\label{fig5}
  \vspace{-1em}
\end{figure}

In this section, we evaluate the described performance indicators for the proposed HAPS+MBS offloading algorithm and compare it with benchmark methods.
\\\indent Figure \ref{fig4} analyzes the impact of increasing traffic demand on network energy efficiency for three different systems comprising 10, 30, and 50 SBSs, respectively. Specifically, for the case of 10 SBSs, we evaluate the efficiency of our proposed method by comparing it with ES method, given the exponential complexity of ES for higher $ {s} $. The results indicate a close performance between the proposed approach and ES in terms of efficiency, validating its effectiveness. The figure reveals two significant observations. Firstly, across all methods, there is a decrease in network energy efficiency as $ {s} $ increases. This decline can be attributed to the overall power consumption rise resulting from the larger $ {s} $. Secondly, as the total traffic demand increases, the performance of the MBS offloading method deteriorates, primarily due to the limited capacity available for offloading SBSs. In contrast, the HAPS+MBS offloading method stands out by demonstrating its capacity to handle the increasing traffic demand more effectively than other methods.
\\\indent Figure \ref{fig3} illustrates the network's total and grid power consumption for different numbers of SBSs. As $ {s} $ increases, all three methods show an upward trend in power consumption. However, this increase in $ {s} $ allows SBSs to serve more traffic, leading to greater free capacity for both HAPS and MBS. Consequently, a larger number of SBSs can be offloaded to HAPS and MBS, resulting in reduced power consumption compared to the no offloading scenario. The proposed HAPS+MBS offloading algorithm proves to be efficient in optimizing power utilization, especially with the increasing $ {s} $, where it reduced grid power consumption by more than ${30\%}$ for ${s = 70}$ compared to the no offloading case.
\\\indent Figure \ref{fig5} illustrates the capacity utilization based on the number of SBSs. Initially, all evaluated methods demonstrate a decline in performance until a turning point at ${s=45}$. 
The primary reason for this decline is the inadequate number of SBSs, resulting in the MBS operating at full capacity and being unable to offload any SBSs. As a result, the MBS offloading method performs similarly to the no offloading approach, while the HAPS becomes the primary contributor in the HAPS+MBS offloading approach.
Beyond this turning point, increasing $ {s} $ reduces MBS congestion and allows for offloading of SBSs based on the predefined algorithm conditions. Consequently, the performance of the MBS offloading method stabilizes when the number of active SBSs remains constant. On the other hand, the performance of the HAPS+MBS offloading method improves as the number of active SBSs decreases, benefiting from the increased available capacity in the HAPS and MBS. Notably, our method achieved a remarkable improvement in capacity utilization, surpassing MBS offloading by at least 20\% for any value of $ {s} $. Furthermore, we used ES method as a benchmark for just the number of SBSs from 10 to 15 due to its increasing complexity. Our simulations indicate that the proposed method has a very close performance to the optimal performance of ES.

\section{Conclusion}
In this article, we introduced a novel HAPS-enabled cell switching method that enhances the energy sustainability of terrestrial networks. Our method significantly improves state vector selection in dense integrated HAPS-terrestrial networks (vHetNet), resulting in reduced network energy consumption while maintaining QoS. Through extensive evaluations and comparisons with benchmark methods, we have demonstrated the superior performance of our proposed algorithm in terms of energy efficiency, network power consumption, and capacity utilization. 

\ifCLASSOPTIONcaptionsoff
  \newpage
\fi
\bibliographystyle{IEEEtran}
\bibliography{IEEEabrv,Bibliography}
\vfill
\end{document}